**Title:**

Transport of impact ejecta from Mars to its moons as a means to reveal Martian history

**Authors:**

*Ryuki Hyodo (hyodo.ryuki@jaxa.jp)[1,2], Kosuke Kurosawa[3], Hidenori Genda[2], Tomohiro Usui[1], Kazuhisa Fujita[1]

**Affiliations:**

[1] ISAS, JAXA, Sagamihara, Japan
[2] Earth-Life Science Institute, Tokyo Institute of Technology, Tokyo 152-8550, Japan
[3] Planetary Exploration Research Center, Chiba Institute of Technology, Chiba, Japan

**Abstract:**

Throughout the history of the solar system, Mars has experienced continuous asteroidal impacts. These impacts have produced impact-generated Mars ejecta, and a fraction of this debris is delivered to Earth as Martian meteorites. Another fraction of the ejecta is delivered to the moons of Mars, Phobos and Deimos. Here, we studied the amount and condition of recent delivery of impact ejecta from Mars to its moons. Using state-of-the-art numerical approaches, we report, for the first time, that materials delivered from Mars to its moons are physically and chemically different from the Martian meteorites, which are all igneous rocks with a limited range of ages. We show that Mars ejecta mixed in the regolith of its moons potentially covers all its geological eras and consists of all types of rocks, from sedimentary to igneous. A Martian moons sample-return mission will bring such materials back to Earth, and the samples will provide a wealth of "time-resolved" geochemical information about the evolution of Martian surface environments.

**Main Text:**

Mars is always compared to Earth because Mars is recognized as a "paleo-habitable" planet. It is an important planet for understanding the evolution of habitable planet(s) in the solar system because of its location near the outer edge of the habitable zone. Throughout the history of the solar system, Mars has experienced numerous asteroidal impacts. These impacts have produced impact-generated Mars ejecta, a fraction of which has been delivered to Earth as Martian meteorites. Another fraction of the ejecta has been delivered to the moons of Mars, Phobos and Deimos (Supplementary Materials, S1).

Phobos and Deimos, the two small moons of Mars, are the target bodies of the Japan Aerospace eXploration Agency (JAXA) sample return mission Martian Moons eXploration (MMX). MMX plans to collect surface material from Phobos and return samples to Earth[1]. Therefore, at this time, before the spacecraft is launched, detailed study of the surface materials on Phobos (potential sample material) is required to maximize the scientific results of the MMX mission.

The origin of the Martian moons is currently still a controversial question. Historically, they were considered captured asteroids originating from the outside of the main belt region; this was based on their spectral properties, which shared a resemblance with D-type asteroids. However, explaining their circular and coplanar orbital properties based on their dynamic aspects[2,3] seems to be a challenge. A giant impact scenario, where a basin-forming impact produced debris from which the Martian moons accreted, would explain the moons' orbital properties naturally[4,5,5,7,8,9,10].

The endogenous bulk material properties of the moons depend on their origins[7]; sampling these materials would provide definitive information[11]. In the capture scenario, the returned sample would be analogous to a type of chondrite[11]. In the giant impact scenario, the bulk materials would be a mixture of Martian and impactor materials that experienced a high temperature phase[7,12,13]. In addition, regardless of the formation scenario, ejecta from Mars produced by the frequent impacts on it (Supplementary Information S1-5) and materials delivered by the natural influx of asteroids (Supplementary Information S6) are mixed within the surface regoliths of the moons. Although such Mars ejecta are not explicitly resolved in previous observations and spectral images of Phobos and Deimos due to instrument limitations, such Mars ejecta can be collected by a sample return mission, as we show below.

In this work, numerical simulations were performed to study the amount and condition of a recent delivery of impact ejecta from Mars to its moons (Supplementary Materials, Materials and Methods). Firstly, impact simulations under a variety of impact conditions expected for asteroidal impacts on Mars were performed to obtain datasets of positions and velocities of the impact debris immediately after impact. The resultant datasets were then used in 30,000 Monte Carlo runs of impact bombardment on Mars, which include the long-term orbital evolution of the impact debris, to assess the quantitative likelihood of delivery to the Martian moon(s) (Supplementary Materials, S2). The impact location on the surface of Mars, orbital phase of the moons were fully randomized (hereafter, "random" case). Other impact parameters, such as impact velocity and impactor mass, were chosen from the expected impactor distribution.

In addition, because the orbit of Phobos is shrinking progressively and the recent decay of the moon's orbit is significant as the tidal evolution strongly depends on the distance to Mars, the delivery from the five largest recent (<10 Myr) craters, Mojave, Tooting, McMurdo, Corinto, and Zunil[14,15,16,17], were also studied selectively by 10,000 Monte Carlo runs for each crater (hereafter, "selected" case). In these specific cases, the impact location on the Martian surface and the impact energy were fixed as that of the current crater position and that capable of producing the observed crater diameter, respectively (Supplementary Materials, Materials and Methods).

**Results**
**Mass transfer from Mars to its moons.**
The mass transferred from Mars to Phobos obtained from the Monte Carlo runs ("random" and "selected" cases) is displayed in Fig. 1 (see also Supplementary Materials, S2). Impacts that produce a larger crater eject a larger amount of Martian debris, and a larger amount of the impact debris reaches Phobos. The dispersion of the delivered mass is mainly attributable to the orbital phase of Phobos in association with the location of the impact (Supplementary Materials, S1). If Phobos is nearly above the point of impact at the time of impact, more efficient delivery of impact debris to the moon(s) occurs, and vice versa.

Recent net mass transfer from Mars to Phobos was calculated as follows. First, an empirical fitting to the Monte Carlo runs of the "random" case was derived from the data obtained at crater sizes between 2 and 300 km (solid line in Fig. 1; Supplementary Materials, S2). Then, using the crater isochrones optimized for Mars[18] and the obtained empirical fitting, the amount of Martian debris delivered during the real impact history of Mars was integrated (Supplementary Materials, S2). Our fitting was applied to craters whose total number is more than 10 in Hartmann isochrons[18], e.g., crater diameter smaller than ~100 km within 500 Myr, to find a statistical value (hereafter, the "statistical" value). In contrast, when the expected number of craters in the isochrones is less than 10, we considered these craters as a stochastic event of a few stochastic large impacts and thus an additional delivery of impact ejecta to the moon was expected to occur stochastically (hereafter, the "stochastic" value).

Assuming that the Martian materials delivered are mixed homogeneously within a 1 m-depth of Phobos regolith (Supplementary Materials, S5) and that the orbit of Phobos remains at the current orbital distance from Mars, our Monte Carlo runs combined with the Martian isochrones[18] found that the amount of Mars ejecta delivered to Phobos within 500 Myr is ~4.9 × $10^9$ kg, or ~1700 ppm in Phobos regolith as a median value (Phobos regolith is assumed to have a density of 1.876 g/cm$^3$ and a mass calculated as ~2.9 × $10^{12}$ kg)[19]. Considering the orbit of Phobos at 500 Myr ago, the cross section becomes about one-fifth[20]. Using this conservative value of ~1/5, at least an ~340 ppm Mars fraction is estimated to have been transported to Phobos regolith during the last 500 Myr (the "statistical" value). In addition, by considering a single large impact event that forms a $D \sim 260$ km, a few stochastic large impacts can add an additional ~1.1 × $10^{10}$ kg × 1/5 = ~2.2 × $10^9$ kg or ~760 ppm (the "stochastic" value). The crater diameter of 260 km was chosen because such a large crater should have been produced at least once within the past 500 Myr, as inferred from the crater isochrones (Supplementary Materials, S2). The total mass delivered to Phobos, including the stochastic value, is ~10 times larger than that estimated previously[20], and this significant update in our work is mainly due to the realistic, direct impact simulations in our work compared to the simple analytical model used in previous studies (Supplementary Materials, S2 and S3), e.g., the previous work considered only head-on collisions, but more probable impacts of 45 degrees produce a much larger amount of ejecta (Fig. 2).

When considering the recent delivery from the five largest craters within the last 10 Myr, the tidal orbital evolution during the past 10 Myr is negligible and Phobos is assumed to have its current orbital distance from Mars. In this case, Mojave Crater makes the largest contribution, and the Mojave-forming event delivered ~9.6 × $10^7$ kg, or ~33 ppm as the median value, of Martian material to Phobos (Fig. 1), which significantly updates the estimated recently delivered mass to Phobos, e.g., about 100 times larger than that reported previously[21]. This difference occurred because the age of Mojave Crater was not well studied at the time the previous paper was published. The expected values of mass delivered from the other large craters were also updated, i.e., ~10 times larger than in previous work (Supplementary Materials, S2 and S3), due to the more realistic treatment of our impact simulations (Fig. 1).

**Physical and chemical properties of Martian materials on its moons.**
The plan for MMX is to collect samples of >10 g of Phobos regolith[22]. Assuming a typical cubic grain diameter for the Mars fraction of 0.3 mm and a density of 3 g/cm$^3$ (the grain mass is ~10$^{-4}$ g), our results indicate that at least ~34 Martian grains (~340 ppm as the "statistical" value) would exist in the returned sample. In addition, another ~76 Martian grains from stochastic large impacts are also potentially available ("stochastic" value of 760 ppm). Moreover, our Monte Carlo runs together with the direct impact simulations indicate that materials delivered from Mars to Phobos and Deimos are physically and chemically different from the Martian meteorites recovered on Earth. We show that the Mars ejecta delivered to Phobos includes less shocked (< 5 GPa) and more fragile materials than those of the Martian meteorites[23] (> 5 GPa) (Fig. 2). The most frequent impacts were oblique; these impacts (with a collision angle of $\theta < 90°$) provide Phobos with a larger amount of less-shocked materials than those provided by head-on collisions that are less frequent (with $\theta = 90°$; supplementary materials, S4).

**Scientific values of Martian materials on its moons.**

Throughout the history of Mars, the delivery of Mars ejecta to Phobos has not been by a single event but rather by numerous events from random locations on the Martian surface, potentially covering the seven geologic eras: pre-Noachian, Noachian (early, middle, & late), Hesperian (early & late), Amazonian (Fig. 3). This suggests that the regolith on Phobos could provide unprecedented information covering the entire history of Mars, cf., Martian meteorites are mostly young, <1.3 billion-year-old igneous rocks[23]. Furthermore, each Martian grain in the regolith on Phobos with a typical grain diameter of 300 μm is expected to contain more than one "chronometer" mineral (e.g., zircon, baddeleyite, or Ca-phosphate for U–Pb dating) in the case that the typical size (<30 μm) and abundance (~0.1–1 %) of chronometer minerals are comparable to those of Martian meteorites[24]. Thus, the >34 Martian grains returned by MMX would provide a wealth of "time-resolved" information on the evolution of Martian surface environments (Fig. 3 top panel) because these grains would have been transported randomly from any of the seven geologic units.

Some of these Martian grains might represent relatively fragile sedimentary materials due to the low shock pressures (Fig. 2). Martian sediments contain aqueous alteration phases such as clays, carbonates, sulfates, and chlorites[25] (Fig. 3 bottom panel). As these alteration phases formed by interaction with surface/subsurface fluids, their geochemical and mineralogical data would provide information about the Martian atmosphere and hydrosphere at the time of their formation. For example, the evolution of surface water/ice and its interaction with the atmosphere can be traced by measurements of hydrogen isotope ratios (D/H: deuterium/hydrogen) in Martian sedimentary grains (Fig. 3 middle panel). Hydrogen is a major component of water ($H_2O$), and its isotopes fractionate significantly during hydrological cycling between the atmosphere, surface waters, ground ice, and the polar ice caps[26]. However, the current D/H dataset is limited by the telescopic measurements that help constrain the hydrogen isotopic compositions of the present-day atmosphere and by the analyses of Martian meteorites that constitute a chronologically biased sampling of the Martian crust[22]. Moreover, the least shocked and least heated Martian grains might preserve a record of remanent magnetization (Fig. 3 middle panel). Thus, the Martian grains in Phobos regolith could fill a knowledge gap and provide a wealth of "time-resolved" geochemical and geophysical information that would facilitate understanding of the co-evolution of the Martian atmosphere, hydrosphere, cryosphere, and magnetosphere.

The < 5 million-year-old crater formation, Mojave-forming event, may deliver additional Martian materials, excavated at a specific time and from a specific site on the Martian surface, to Phobos (with ~3 Martian grains as the median value in the sample of 10 g). It has been suggested that Martian meteorites known as the shergottites originate from the Mojave crater[17]. The returned samples may contain an analog to the shergottites, which will be also used to constrain the putative link between the origin of the Martian meteorites and crater formation.

**Discussion**

Mars rovers, such as Curiosity and Mars 2020, are ongoing and planned[27,28]. They are designed to acquire specific detailed information about Martian geology near their landing sites. A series of future Mars sample-return (MSR) missions are also planned for a specific crater, Jezero Crater[28]. Their biggest advantage is that they investigate a specific local geological context in great detail. At the same time, their disadvantages are that the data obtained are necessarily limited to local information and that they presumably access mostly old rocks near the surface. Martian meteorites provide us with data-rich information, but they are limited by a lack of geological context as the point of origin is uncertain and they are all igneous rocks within a limited range of ages. Samples from Phobos would be relatively small (~10 g) compared to Martian meteorites, and their original geological context would be unknown in contrast to MSR and in situ measurements by rovers. However, they could potentially cover all geological eras and contain all types of rocks from sedimentary to igneous. Hence, together with the other detailed and specific information obtained from MSR, Mars rovers, and Martian meteorites, the randomized nature of the particles from Mars on Phobos can play the complementary roles of revealing the time-resolved geochemistry of Mars (see Fig. 3) and extending the scientific concept of Martian moons from simply "Martian moon science" to "Mars system science".

**_References:_**

1. Kawakatsu, Y. *et al.* Mission concept of Martian Moons eXploration (MMX). In *68th International Astronautical Congress (IAC)*, IAC-17-A3.3A.5. (2017).
2. Burns, J.A. Contradictory clues as to the origin of the Martian moons. H.H. Kieffer, B.M. Jakosky, C.W. Snyder, & M.S. Matthews (Eds.), Mars (pp. 1283-1301). Tuscon, Arizona, USA: University of Arizona Press (1992).
3. Murchie, S.L. *et al.* Color heterogeneity of the surface of Phobos: Relationships of geological features and comparison to meteorite analogs. *Journal of Geophysical Research,* **96**, 5925-5945 (1991).
4. Craddock, R.A. Are Phobos and Deimos the result of a giant impact? *Icarus* **211**, 1150-1161 (2011).
5. Citron, R.I., Genda, S., & Ida, S. Formation of Phobos and Deimos via a giant impact. *Icarus,* **252***,* 334-338 (2015).
6. Rosenblatt, P. *et al.* Accretion of Phobos and Deimos in an extended debris disc stirred by transient moons. *Nature Geo.* **9**, 581 (2016).
7. Hyodo, R., Genda, H., Charnoz, S., & Rosenblatt, P. On the Impact Origin of Phobos and Deimos. I. Thermodynamic and Physical Aspects. *Astrophys. J.* **845**, 125 (2017).
8. Hesselbrock, A., & Minton, D. A. An ongoing satellite-ring cycle of Mars and the origins of Phobos and Deimos. *Nature Geo.* **10**, 266 (2017).
9. Hyodo, R., Rosenblatt, P., Genda, H., & Charnoz, S. On the Impact Origin of Phobos and Deimos. II. True Polar Wander and Disk Evolution. *Astrophys. J.* **851**, 122 (2017).
10. Canup, R., & Salmon, J. Origin of Phobos and Deimos by the impact of a Vesta-to-Ceres sized body with Mars. *Sci. Adv.* **4**, eaar6887 (2018).
11. Usui, T. et al. Martian Moons Exploration: The Importance of Phobos Sample Return for Understanding the Mars-Moon System. *LPI Contribution* No. 2132, id.2388 (2019).
12. Pignatale, F. *et al.* On the Impact Origin of Phobos and Deimos. III. resulting composition from different impactors. *Astrophys. J.* **853**, 118 (2018).
13. Hyodo, R., Genda, H., Charnoz, S., Pignatale, F. & Rosenblatt, P. On the Impact Origin of Phobos and Deimos. IV. Volatile depletion. *Astrophys. J.* **860**, 150 (2018).
14. Golombek, M., Bloom, C., Wigton, N. & Warner, N. Constraints on the age of Corinto crater from mapping secondaries in Elysium planitia on Mars. *LPS XXXXV* 1470 (2014).

15.  Hartmann, W.K., Quantin, C., Werner, S.C. & Popova, O. Do young martian ray craters have ages consistent with the crater count system? *Icarus* **208**, 621–635 (2010).
16.  Malin, M.C., Edgett, K., Posiolova, L., McColley, S. & Noe Dobrea, E. Present impact cratering rate and the contemporary gully activity on Mars: Results of the Mars Global Surveyor extended mission. *Science* **314**, 1557-1573 (2006).
17.  Werner, S.C., Ody, A. & Poulet, F. The source crater of Martian Shergottite meteorites. *Science* **343**, 1343-1346 (2014).
18.  Hartmann, W.K. Martian cratering 8: Isochron refinement and the chronology of Mars. *Icarus* **174**, 294-320 (2005).
*19.*  Andert, T. P. *et al.* Precise mass determination and the nature of Phobos. *Geophys. Res. Lett.* **37**, L09202 (2010).
20.  Chappaz, L., Melosh, H.J., Vaquero, M. & Howell, K.C. Transfer of impact ejecta material from the surface of Mars to Phobos and Deimos. *Astrobiology* **13**, 963-980 (2013).
21.  Ramsley, K.R. & Head, J.W. Mars impact ejecta in the regolith of Phobos: Bulk concentration and distribution. *Planetary and Space Science* **87**, 115-129 (2013).
22.  Usui, T. Hydrogen reservoirs in Mars as revealed by martian meteorites, in *Volatiles in the Martian Crust*, ed. by J. Filiberto, S.S. Schwenzer (2019).
23.  Nyquist L.E. *et al.* Ages and Geologic Histories of Martian Meteorites. In: Kallenbach R., Geiss J., Hartmann W.K. (eds) Chronology and Evolution of Mars. *Space Sciences Series of ISSI*, vol 12. Springer, Dordrecht (2001).
24.  McSween H.Y. & Treiman A.H. Martian meteorites. In: Papike J.J. (ed.) *Planetary Materials. Reviews* in Mineralogy, vol. 36, pp. 6–1–6-53. Washington, DC: Mineralogical Society of America (1998).
25.  Ehlmann, B.L. & Edwards, C.S. *Mineralogy of the Martian Surface. AREPS* **42**, 291 (2014).
26.  Villanueva, G. L. *et al.* Strong water isotopic anomalies in the Martian atmosphere: Probing current and ancient reservoirs. *Science* **348** (6231), 218-221 (2015).
27.  Grotzinger J. P. *et al*. A Habitable Fluvio-Lacustrine Environment at Yellowknife Bay, Gale Crater, Mars (NewYork, N.Y.). *Science* **343**, 1242777 (2014).
28.  Beaty, D.W. *et al*. The potential science and engineering value of samples delivered to Earth by Mars sample return. *Meteoritics & Planetary Science* **54**, S3-S152 (2019).
29.  Lillis, R.J., Frey, H.V. & Manga, M. Rapid decrease in Martian crustal magnetization in the Noachian era: Implications for the dynamo and climate of early Mars. *Geophys. Res. Lett.* **35**, L14203 (2008).
30.  Kurokawa, H. *et al*. Evolution of water reservoirs on Mars: Constraints from hydrogen isotopes in martian meteorites. *Earth Planet. Sci. Lett.* **394**, 179-185 (2014).
31.  Hartmann, W.K. & Neukum, G. Cratering chronology and the evolution of Mars. *Space Sci. Rev.* **96** (1/4), 165-194 (2001).

## *Acknowledgments:*

**Funding:** R.H. acknowledges the financial support of JSPS Grants-in-Aid (JP17J01269, 18K13600). K.K. is supported by JSPS KAKENHI grant numbers 17H01176, 17H02990, 17H01175, 17K18812, 18HH04464, 19H00726 and by the Astrobiology Center of the National Institute of Natural Sciences, NINS (AB301018). H.G. is supported by MEXT KAKENHI Grant No. JP17H06457. T.U. acknowledges the financial support of MEXT KAKENHI Grant No. JP17H06459. H.G. and K.K. is supported by JSPS Kakenhi Grant No. JP17H02990. We would like to thank Editage (www.editage.com) for English language editing. **Author contributions:** R.H., K.K., H.G. and K.F. conceived of the project, designed the study; R.H. wrote the paper; R.H and T.U. wrote the discussion about the future Mars explorations; R.H. and K.K. performed the Monte-Carlo orbital simulations and analyzed the data; H.G. performed impact simulations. We thank Henry Jay Melosh and an anonymous referee for their constructive comments that helped greatly improve the manuscript. **Competing interests:** The authors declare no competing interests. **Data and materials availability:** All data to understand and assess the conclusions of this research are available in the main text and supplementary materials.

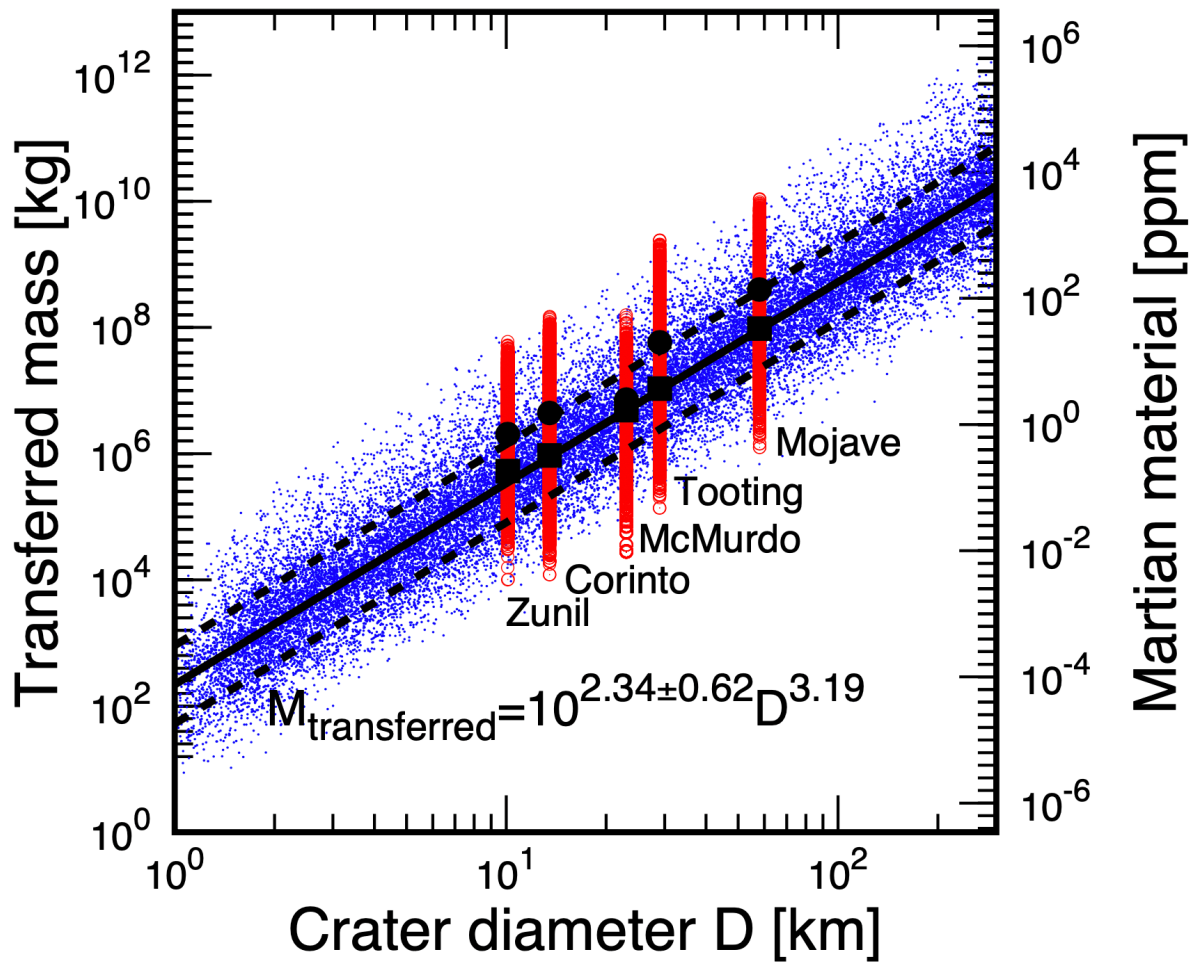

**Fig. 1. The mass transferred from Mars to the Martian moon Phobos.** The blue points show the results of 30,000 impacts of fully randomized cases. The red points show those of the five largest recent crater-forming events. The large squares and large circles represent the median and mean values of the cases of largest crater-forming events, respectively. The solid line represents the median values of the fully randomized cases fitted using the data from $D$ = 2–300 km ($M_{\text{transferred}} = 10^{2.34}D^{3.19}$). The dashed lines represent the dispersions in which ~70% of the transferred mass is covered within the two dashed lines ($M_{\text{transferred}} = 10^{2.34+0.62}D^{3.19}$ and $M_{\text{transferred}} = 10^{2.34-0.62}D^{3.19}$). *Y*-axis on the right side of the panel shows the corresponding fraction of Martian materials assuming they are mixed homogeneously within a 1 m-depth of Phobos regolith.

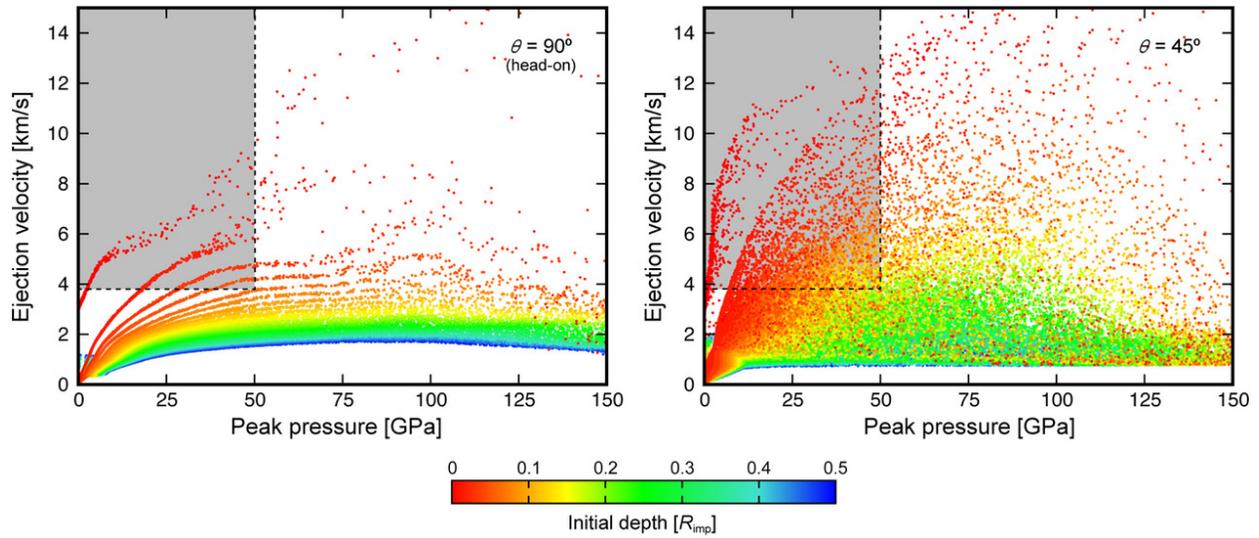

**Fig. 2. Ejection velocity as a function of the peak pressure experienced during the impact and ejection process (results of typical impact velocity of 12 km/s to Mars).** Shadowed regions indicate ejecta capable of reaching Phobos ($V_{eje} > 3.8$ km/s). Compared to Martian meteorites ($P_{pk} > 5$ GPa (5–50 GPa) and $V_{eje} > 5$ km/s), ejecta that can reach Phobos is less shocked ($P_{pk} < 5$ GPa) and includes more primitive chronometer grains that can trace the time-evolution of Mars. The number of plotted particles in the panels is reduced by 1/30 from that used in the calculations to avoid too many points.

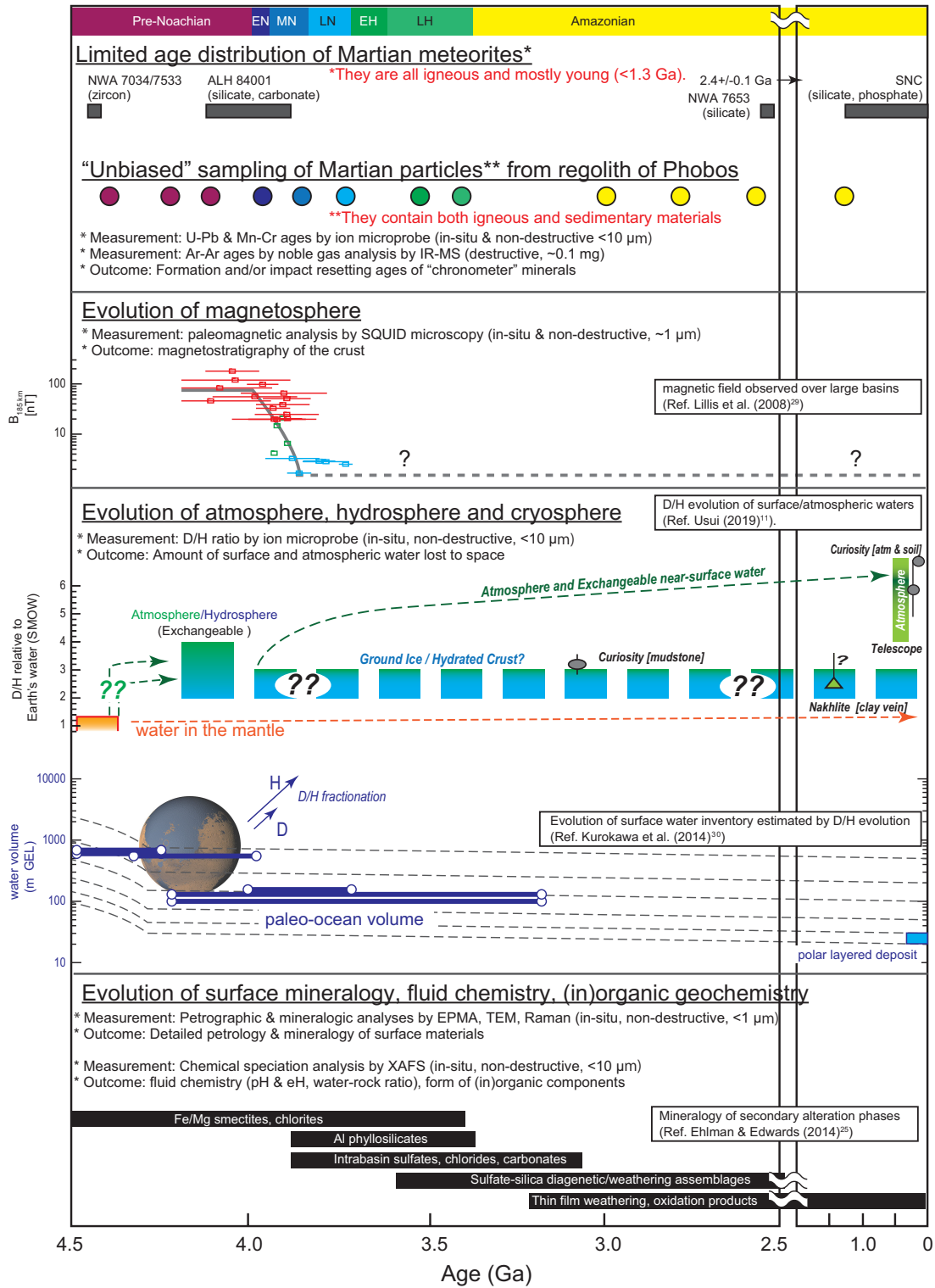

**Fig. 3. Formation age distributions of Martian meteorites and Martian particles in the regolith of Phobos compared with the evolution of magnetic activity, D/H ratio of hydrosphere/cryosphere, surface water inventory, surface mineralogy, and key stratigraphies created utilizing the Hartmann and Neukum (2001) chronology model.** X-axis is time in the unit of Ga.